\documentclass[conference]{IEEEtran}
\usepackage{amsmath,graphicx}
\usepackage{bm}
\usepackage{amssymb}
\usepackage{algorithm}
\usepackage{algorithmic}
\usepackage{booktabs}
\usepackage{fancyhdr}
\usepackage[font=footnotesize]{caption}
\usepackage[font=footnotesize]{subcaption}
\usepackage{mathtools}
\usepackage{url}
\usepackage[per-mode=repeated-symbol,input-symbols = ()]{siunitx}
\def\BibTeX{{\rm B\kern-.05em{\sc i\kern-.025em b}\kern-.08em
		T\kern-.1667em\lower.7ex\hbox{E}\kern-.125emX}}

\newcommand{\Mod}[1]{\ \mathrm{mod}\ #1}
\newcommand{\sfrac}[2]{\textstyle \frac{#1}{#2}}

\begin{document}

	\title{Signal Identification and Entrainment for Practical FMCW Radar Spoofing Attacks}
	
	\author{
		\fontsize{10}{12}\selectfont 
		\IEEEauthorblockN{Andrew M. Graff\IEEEauthorrefmark{1}, Todd E. Humphreys\IEEEauthorrefmark{2}}\\
		\IEEEauthorblockA{
			\IEEEauthorrefmark{1}Department of Electrical and Computer Engineering, The University of Texas at Austin\\
			\IEEEauthorrefmark{2}Department of Aerospace Engineering and Engineering Mechanics, The University of Texas at Austin
			\\\{andrewgraff,\ todd.humphreys\}@utexas.edu}
	}
	
	\maketitle
	
	%
	\maketitle
	\begin{abstract}
		This paper proposes a method of passively estimating the parameters of frequency-modulated-continuous-wave (FMCW) radar signals with a wide range of structural parameter values and analyzes how a malicious actor could employ such estimates to track and spoof a target radar. When radars are implemented to support automated driver assistance systems, an intelligent spoofer has the potential to substantially disrupt safe navigation by inducing its target to perceive false objects. Such a spoofer must acquire highly accurate estimates of the target radar's chirp sweep, timing, and frequency parameters while additionally tracking and compensating for time and Doppler shifts due to clock errors and relative movement. This is a difficult task for millimeter-wave radars due to severe Doppler shifts and fast sweep rates, especially when the spoofer uses off-the-shelf FMCW equipment. Algorithms and techniques for acquiring and tracking an FMCW radar are proposed and verified through simulation, which will help guide future decisions on appropriate radar spoofing countermeasures.
		
	\end{abstract}
	\begin{IEEEkeywords}
		FMCW, radar, passive, spoofing, mmWave
	\end{IEEEkeywords}

	\newif\ifpreprint
	\preprinttrue
	
	\ifpreprint
	
	\pagestyle{plain}
	\thispagestyle{fancy}  
	\fancyhf{} 
	\renewcommand{\headrulewidth}{0pt}
	\rfoot{\footnotesize \bf Preprint of the 2023 IEEE 97th Vehicular Technology Conference\\ Florence, Italy, June 20-23, 2023} \lfoot{\footnotesize \bf 979-8-3503-1114-3/23/\$31.00~\copyright2023 IEEE }
	
	\else
	
	\thispagestyle{empty}
	\pagestyle{empty}
	
	\fi
	

	%
	\section{Introduction}
	\label{sec:intro}
	
	Frequency-modulated continuous-wave (FMCW) radars operate by transmitting a sequence of multiple linear chirps called a frame. After reflecting off an object, the received signal is mixed with a replica of the transmitted chirp sequence, resulting in a sinusoid whose frequency indicates the range to the reflecting object. By examining the phase shifts of this sinusoid across multiple chirps, a radar can estimate the object's relative velocity. Furthermore, an FMCW radar can estimate an object's direction by examining the phase shifts across elements in its antenna array.
	
	Millimeter-wave (mmWave) radars have significantly improved radar precision \cite{hasch2012millimeter}, providing benefits over alternative sensors for use in automated driver assistance systems (ADAS) and autonomous vehicle (AV) platforms \cite{sun2020mimo,patole2017automotive}. Radars are currently deployed in ADAS for adaptive cruise control, forward collision avoidance, and lane-change assist, among others \cite{engels2017advances}. In more advanced systems, these radars may be used for imaging by providing point cloud data to computer vision systems within AVs \cite{sun2020mimo}. Beyond the automotive industry, these radars can provide precise sensing for urban air mobility applications \cite{lies2021longRange}. With an ever increasing number of automated systems relying on mmWave FMCW radar, it is crucial that these sensors remain secure against potential threats.
	
	Since typical fast-chirp mmWave FMCW radars use an intermediate frequency (IF) bandwidth in the 10s of \SI{}{\mega\hertz} and chirp slopes in the 10s of \SI{}{\mega\hertz\per\micro\second}, these radars only sample a thin sliver of the time-frequency spectrum at any instant, making it unlikely that persistent interference will appear in the band of sensitivity and manifest as a false reflecting object \cite{D12mosarim2010,D1mosarim2010}. However, existing fast-chirp processing is susceptible to deliberate spoofing. By controlling the time-of-arrival and frequency offset of the spoofing signal, an attacker can induce the target receiver to perceive fake objects at any arbitrary range and velocity. Since FMCW radars are the most widely used radars in automotive vehicles \cite{alland2019interference}, such spoofing could have widespread ramifications.
	
	Previous studies on radar spoofing have focused directly on the mechanics of a spoofing attack, demonstrating FMCW spoofing attacks with commercial-off-the-shelf (COTS) devices \cite{komissarov2021spoofing,miura2019low,ordean2022millimeter}, with custom boards \cite{nallabolu2021frequency}, and with an AV testbed \cite{sun2021control}. However, this work assumes that the spoofer already knows the target radar's waveform, which is unlikely in a real-world scenario. Furthermore, the advanced spoofing attack in \cite{sun2021control} was carried out using high-end test equipment, which may not be representative of the most common radar spoofing threats. Separate from spoofing literature, the related problem of FMCW parameter estimation and synchronization has been examined. Gardill et al. proposed a method of finding an unknown FMCW signal by analyzing the time-frequency spectrum of interference when mixed with a local fast-sweep-rate FMCW signal \cite{gardill2019situ}. Their study was then extended to demonstrate how such a tactic could be used to first estimate signal parameters, switch the local mixer to a CW signal to obtain precise timing, and then switch the local mixer to a time-aligned replica of the transmitted signal \cite{gardill2020approach}. While these studies did not directly address FMCW spoofing, they show that synchronization schemes are practical and that issues such as timing jitter can be accounted for. However, the papers do not propose a rigorous method of chirp parameter estimation and tracking. Other work has focused purely on synchronizing FMCW systems when the signal shape is known \cite{roehr2007radiomethod,lampel2020system}. Work on counter-adversarial tracking has also analyzed the estimation and countering of cognitive radars \cite{krishnamurthy2021adversarial}.
	
	Prior work has studied passive FMCW parameter estimation and FMCW spoofing attacks but falls short of detailing a comprehensive attack that can be carried out using COTS radars without prior signal knowledge. This paper addresses these gaps so future work can evaluate effective countermeasures.
	
	\section{System Model}
	\label{sec:model}
	
	Consider an environment where there is a set of radars $\mathcal{I}$, all with standard FMCW architectures, and each having one or more transmit and receive antennas. Each radar has a signal generator that creates linear frequency chirps. A sequence of these chirps, called a frame, is generated and processed coherently. The number of chirps depends on the desired Doppler resolution and can range between tens and hundreds of chirps per frame. Each chirp signal is sent to one or more transmit antennas and to a mixer that takes its other input from the receive antennas. Let the transmitted signal of the $k$th chirp in the $i$th radar's frame be denoted as $s_{ik}(t)$, where $i \in \mathcal{I}$ and $k \in K_{i}$, with $K_{i}$ being the set of chirps in one frame from radar $i$. The $k$th chirp has a start frequency $f_{ik}$, a chirp slope $\beta_{ik}$, a phase offset $\phi_{ik}$, and a duration $T_{ik}$. A baseband model of the $i$th radar's $k$th chirp is
	\begin{align}
		c_{ik}(t) = \begin{cases} 
			\exp{\left(j2\pi \left(f_{ik}t+ \frac{\beta_{ik} t^2}{2}\right) + j\phi_{ik}\right)} & t \in [0, T_{ik})\\
			0 & \text{otherwise}\ .
		\end{cases}
	\end{align}
	
	This chirp signal is sent to one or more transmit antennas with some complex weighting. Assume that radar $i$ has $N_{\text{TX},i}$ transmit antennas and $N_{\text{RX},i}$ receive antennas. Let the vector of transmit antenna weights for chirp $k$ on radar $i$ be $\bm{w}_{ik} \in \mathbb{C}^{N_{\text{TX},i} \times 1}$. For this paper, $\bm{w}_{ik}$ is assumed to be one-hot, meaning that it contains only one nonzero element. However, MIMO radars may use orthogonal codes such as Hadamard codes across multiple antennas, in which case every element of $\bm{w}_{ik}$ would contain $\pm1$ in accordance with the code. Additionally, $\bm{w}_{ik}$ may apply a binary phase-shift keying (BPSK) or quadrature phase-shift keying (QPSK) phase code across each chirp $k$. Each chirp signal is delayed $\Delta t_{ik}$ seconds after the beginning of the frame and weighted by the vector $\bm{w}_{ik}$, creating the transmitted signal
	\begin{align}
		\bm{s}_{i}(t) = \sum_{k \in K} \bm{w}_{ik} c_{ik}(t-\Delta t_{ik}),
		\label{eq:local_sig}
	\end{align}
	
	which propagates along several paths to the receive antennas. Let $\mathcal{R}_{l}^{i}$ be a set of paths from radar $l$ to radar $i$. In the case of monostatic radar, the transmit and receive radars are the same, and this set can be written as $\mathcal{R}_{i}^{i}$. Focusing on the monostatic case, the signal in each path $r \in \mathcal{R}_{i}^{i}$ experiences a propagation delay $\Delta t_r$ and arrives with a receive magnitude $\alpha_{ir}$. Due to the antenna array manifold, relative phase shifts are imparted on the signal depending upon the angles of departure and arrival of each path. The transmitter array phase shift is given by $\bm{a}_{\text{TX},ir} \in \mathbb{C}^{N_{\text{TX},i} \times 1}$ and the receiver array phase shift is given by $\bm{a}_{\text{RX},ir} \in \mathbb{C}^{N_{\text{RX},i} \times 1}$ for the $r$th path. Lastly, the signal propagating along the $r$th path experiences a Doppler shift $f_{\text{D},r}$ arising from the motion of the radar and the reflecting object. This shift is modeled as a multiplication of the signal by $\nu_{r}(t) = \exp{\left(j2\pi f_{\text{D},r}t\right)}$. The received signal across all receive antennas is then given by the vector
	\begin{align}
		\bm{x}_{i}(t) = \sum_{\mathclap{r \in \mathcal{R}_{i}^{i}}} \alpha_{ir}  \bm{a}_{\text{RX},ir}  \bm{a}_{\text{TX},ir}^{\top} \bm{s}_{i}(t-\Delta t_r) \nu_{r}(t).
	\end{align}
	In addition to this monostatic signal, radar $i$ will receive interference from other radars in the environment. The interference signal received at radar $i$ is
	\begin{align}
		\tilde{\bm{x}}_{i}(t) = \sum_{l \in \mathcal{I}\setminus \{i\}}\ \ \sum_{\mathclap{r \in \mathcal{R}_{l}^{i}}} \alpha_{ir}  \bm{a}_{\text{RX},ir}  \bm{a}_{\text{TX},lr}^{\top} \bm{s}_{l}(t-\Delta t_r) \nu_{r}(t).
	\end{align}
	The sum $\bm{x}_{i}(t) + \tilde{\bm{x}}_{i}(t)$ is mixed with the conjugate transpose of radar $i$'s local signal given in (\ref{eq:local_sig}), passed through a lowpass anti-aliasing filter with impulse response $h(t)$, and sampled with period $T_s$, creating the  $N_{\text{RX},i} \times N_{\text{TX},i}$ matrix $\bm{Y}_{i}[n]$:
	\begin{align}
		\bm{Y}_{i}(t) = \left(\bm{x}_{i}(t) + \tilde{\bm{x}}_{i}(t)\right) \bm{s}_{i}^{H}(t).\label{eq:mixing}\\
		\bm{Y}_{i}[n] = \int_{-\infty}^{\infty}h(nT_s - \tau)\bm{Y}_{i}(\tau) \,d\tau.
	\end{align}
	
	Consider the mixed signals originating from a different radar. Let $m_{l\tilde{k}}^{ikr}(t)$ be the resulting signal when the $\tilde{k}$th transmitted chirp from radar $l$ is received along path $r$ by radar $i$ and mixed with its $k$th chirp:
	\begin{align}
		\label{eq:mix_sig_passive}
		m_{l\tilde{k}}^{ikr}&(t) = c_{l\tilde{k}}(t-\Delta{t}_{r}) c_{ik}^{\ast}(t) \nu_{r}(t)\\
		= &\exp{}\Bigl( j2\pi \Bigl( (f_{l\tilde{k}} - f_{ik}) t - \beta_{l\tilde{k}} \Delta{t}_{r} t + f_{\text{D},r} t - f_{l\tilde{k}} \Delta{t}_{r} \nonumber\\
		& + \sfrac{1}{2} (\beta_{l\tilde{k}}-\beta_{ik}) t^2 + \sfrac{1}{2} \beta_{l,\tilde{k}} \Delta{t}_{r}^2 \Bigr)+ j(\phi_{l\tilde{k}}-\phi_{ik})\Bigr)\nonumber.
	\end{align}
	
	This is a linear chirp with a slope of $\beta_{l\tilde{k}}-\beta_{ik}$, a start frequency of $f_{l\tilde{k}} - f_{ik} - \beta_{l\tilde{k}} \Delta{t}_{r} + f_{\text{D},r}$, and a phase shift of $-2\pi f_{l\tilde{k}} \Delta{t}_{r} + \pi \beta_{l\tilde{k}} \Delta{t}_{r}^2 + \phi_{l\tilde{k}}-\phi_{ik}$.
	
	Real-world radar systems are prone to clock errors, and the system model must account for their effects, especially in the case of bistatic propagation where the two radars are unsynchronized. Therefore, the clock frequency and phase drifts are modeled with the two-state model described in \cite[Chapter~9.3]{brown2012introKf}. This model is parameterized by two Allan variance parameters, $h_0$ and $h_{-2}$. Consider two white-noise processes, $F_{i}(t)$ and $G_{i}(t)$, with spectral densities$S_{\text{f},i}  \approx \frac{h_0}{2}$ and $S_{\text{g},i} \approx 2\pi^2h_{-2}$, respectively. The time perceived at radar $i$ is $\tilde{t}_{i}(t)$, a function of the true time $t$:
	\begin{align}
		\tilde{t}_{i}(t) = t + \int_{0}^{t} F_{i}(\tau_f) + \int_{0}^{\tau_f} G_{i}(\tau_g)\,d\tau_g\,d\tau_f.
	\end{align}
	
	Considering only the azimuth dimension, propagation path $r$ has an angle of arrival $\theta_{\text{AOA},r}$ and an angle of departure $\theta_{\text{AOD},r}$. Each radar's antennas have a gain as a function of angle, $G_{\text{RX},i}(\theta)$ and $G_{\text{TX},i}(\theta)$. The signal experiences a path loss of $L_{\text{prop},r}$, which may be one-way for a direct path, or two-way for a reflected path. The reflection in the path has a radar cross section of $\sigma_{\text{RCS},r}$. The total received power is $P_{\text{RX},ir} = P_{\text{TX},l} G_{\text{RX},i}(\theta_{\text{AOA},r}) G_{\text{TX},l}(\theta_{\text{AOD},r}) L_{\text{prop},r}^{-1} \sigma_{\text{RCS}}$; the received magnitude is $\alpha_{ir} = \sqrt{P_{\text{RX},ir}}$.
	
	\section{FMCW Entrainment}
	\label{sec:entrainment}
	
	Consider an instance of the signal model where the set of radars includes one target radar $l$ and one spoofer radar $i$. The spoofer aims to transmit a false replica of the target's signal, causing the target radar to perceive a fake object. However, the spoofer is assumed to have no prior knowledge of the target's signal structure other than it taking the form of a mmWave FMCW signal with repeating frames. To recreate the target signal accurately for spoofing, the spoofer must have an estimate of the parameters in Table \ref{tab:signal_params}. These are the parameters as perceived by the spoofer which may be different from the true target parameters due to ambiguities caused by the clocks, timing, Doppler, and chirp phase shifts. Furthermore, the spoofer is assumed to use COTS FMCW radar equipment. While a wideband mmWave receiver may be capable of easily estimating these parameters after capturing the entire chirp sequence, the small IF bandwidth of COTS FMCW radars complicates this task. Instead, the spoofer must strategically design its mixing signal to collect information about the spectrum and the target signal, only receiving useful samples when the target and mixing signals are close in frequency. The spoofing scheme proposed here offers one possible strategy for estimating the target signal's parameter under these constraints.
	
	\begin{table}[t]
		\centering
		\caption{Required Signal Parameters}
		\begin{tabular}[c]{ll}
			\toprule
			$T_{\text{frame}}$        & Frame interval\\
			$\Delta t_{\text{frame}}$         & Frame timing offset\\
			$T_{k}$        & Duration of the $k$th chirp in the frame\\
			$\Delta t_{k}$        & Start time of the $k$th chirp after the frame start\\
			$\Delta f_{k}$       & Start frequency of the $k$th chirp\\
			$\beta_k$       & Slope of the $k$th chirp\\
			$K$				& Number of chirps in a frame\\
			$\phi_k$     & Phase shift applied to the $k$th chirp \\
			\bottomrule
		\end{tabular}
		\label{tab:signal_params} 
	\end{table}
	
	The proposed FMCW spoofing attack can be divided into three stages: (1) an identification stage where the spoofer samples the spectrum to discover a target radar signal and coarsely estimates the target's parameters, (2) a tracking stage where the spoofer refines its parameter estimates and tracks the signal with greater precision, and (3) a spoofing stage where the spoofer transmits a signal with the intent of appearing as a fake object to the target radar. This paper focuses on the first two stages, which will collectively be called signal entrainment.
	
	\subsection{Identification}
	
\begin{figure}[t]
		\centering
		\includegraphics[width=0.7\linewidth,trim={0 1.5cm 0 2.3cm},clip]{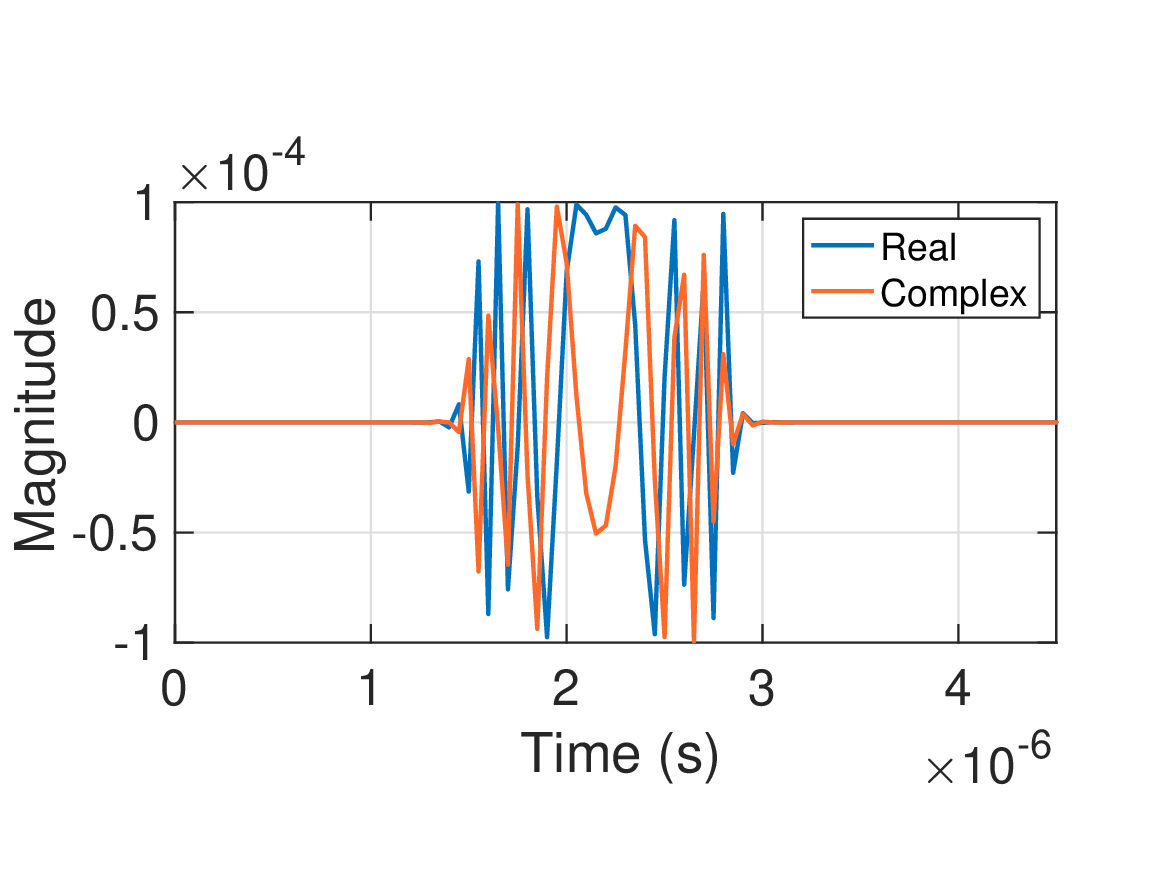}
		\caption{Complex samples of the signal modeled in (\ref{eq:mix_sig_passive}) during an interval in which the mixing signal crosses the received signal in time and frequency, $f_{ik} = f_{\text{tone}}$, and $\beta_{ik}=0$.}
		\label{fig:complex_crossing}
	\end{figure}

	During the identification stage, the spoofer operates passively and attempts to find an FMCW radar to target. Throughout this stage, the spoofer restricts its signal to a constant frequency tone with frequency $f_{\text{tone}}$.
	%
	First, the spoofer attempts to detect the presence of an FMCW signal. From (\ref{eq:mix_sig_passive}), the mixed output of a chirp with a constant frequency tone (i.e., $f_{ik} = f_{\text{tone}}$ and $\beta_{ik}=0$) is a chirp whose instantaneous frequency is $\beta_{l\tilde{k}} (t-\Delta t_r) + f_{l\tilde{k}} - f_{\text{tone}} + f_{\text{D},r}$. Due to the anti-aliasing filter, this signal will be filtered out except for the short time-span when the instaneous frequency is near zero. Fig.~\ref{fig:complex_crossing} visualizes one instance of such a signal. Due to the one-way path loss of the target-spoofer system and beamforming, this signal is easily detectable above the noise floor without any pulse integration. Recall that $\bm{w}_{ik}$ and, consequently, $\bm{s}_{i}(t)$ are assumed to be one-hot. From (\ref{eq:mixing}), this assumption means $\bm{Y}_{i}[n]$ has only one non-zero column. Let this column be written as $\bm{y}_{i}[n]$ with shape $N_{\text{RX},i} \times 1$ in this specific case. Beamforming is performed with an FFT \cite{patole2017automotive,mucci1984comparison} of $\bm{y}_{i}[n]$ across the receive antenna dimension, creating the beamformed vector $\tilde{\bm{y}}_{i}[n]$, whose $b$th element corresponds to the $b$th beamforming bin. Assuming an oversampling factor of $N_{\text{os}}$, $\tilde{\bm{y}}_{i}[n]$ has the shape $N_{\text{os}}N_{\text{RX},i} \times 1$.
	\begin{figure}[t]
		\centering
		\includegraphics[width=0.8\linewidth,page=7]{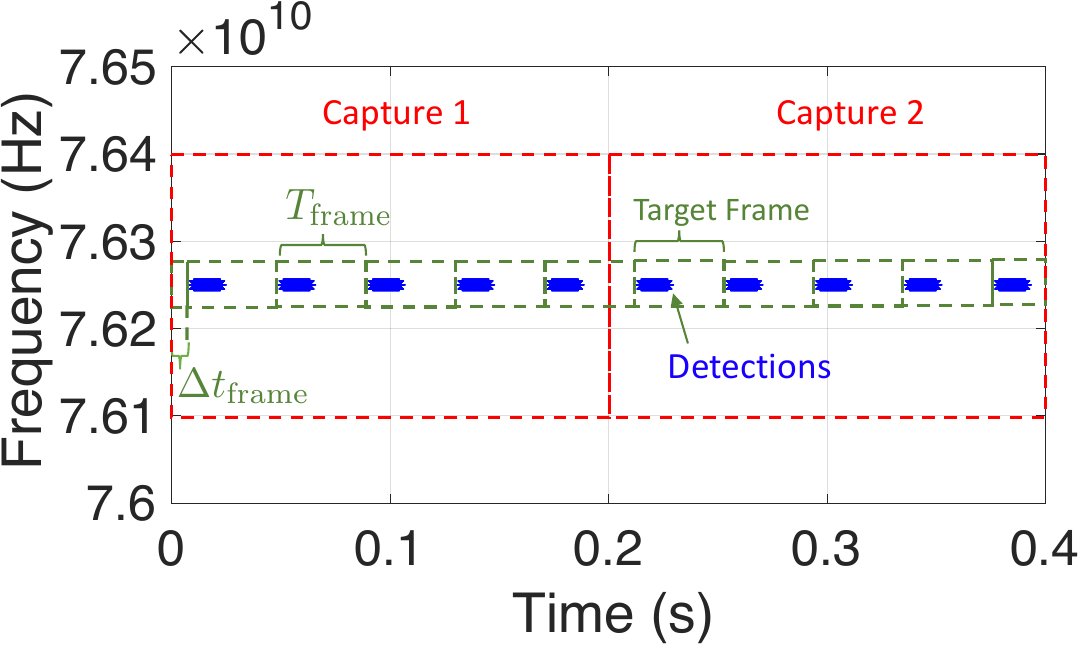}
		\caption{A visualization of the detections obtained during the two captures in the identification stage when mixing with a \SI{76.25}{\giga\hertz} tone. The target signal's frame interval $T_{\text{frame}}$ and frame timing offset $\Delta t_{\text{frame}}$ are highlighted.}
		\label{fig:params_frame}
	\end{figure}
	\begin{figure}[t]
		\centering
		\includegraphics[width=0.8\linewidth,page=8]{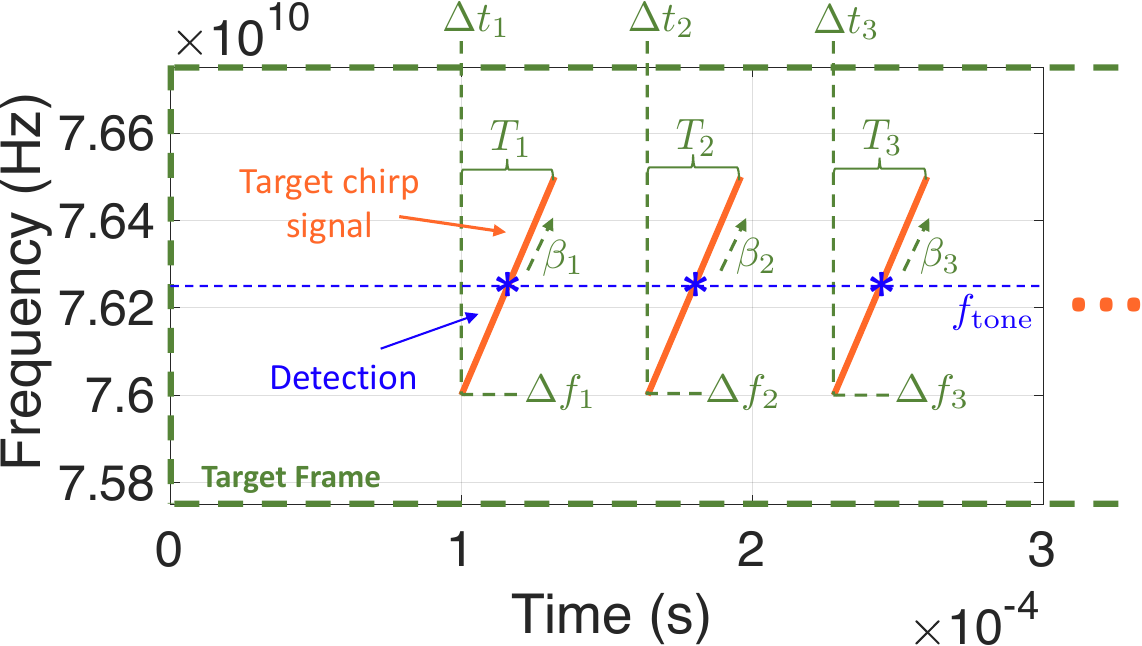}
		\caption{A visualization of the detections obtained during a capture in the identification stage when mixing with a \SI{76.25}{\giga\hertz} tone. The target signal's chirps are shown, highlighting each chirp's start time $\Delta t_k$, start frequency $\Delta f_k$, duration $T_k$, and slope $\beta_k$.}
		\label{fig:params_chirp}
	\end{figure}
	
	First, the spoofer determines which sample index $n$ and beamforming bin $b$ contain the maximum power: $\{\hat{n}_{\text{max}},\hat{b}_{\text{max}}\} = \operatorname*{argmax}_{n,b} |\left(\tilde{\bm{y}}_{i}[n]\right)_b|^2$. The spoofer then uses a constant false alarm rate (CFAR) detector \cite[Chapter~6.5]{radarsp_richards} on the power-maximizing beamformed signal $(\tilde{\bm{y}}_{i}[n])_{\hat{b}_{\text{max}}}$ to create a set of detections $\mathcal{D}_{\text{sample}}$ containing sample indices and frequencies $(\hat{n},\hat{f})$. The frequencies come from the known frequency of the spoofer's mixer, $f_{\text{tone}}$. The set of detections $\mathcal{D}_{\text{sample}}$ is consolidated by replacing groups of neighboring sample indices $\hat{n}$ with a single detection at the center sample of each group. Lastly, a new set $\mathcal{D}$ is created where $	\mathcal{D} = \{(nT_{s},f)\ :\ (n,f)\in\mathcal{D}_{\text{sample}}\}$, converting the sample indices to times.
	
	Now, the spoofer estimates the frame interval $T_{\text{frame}}$ and frame timing offset $\Delta t_{\text{frame}}$, which are defined graphically as shown in Fig.~\ref{fig:params_frame}. The spoofer collects two captures of samples while mixing the received signal with a constant frequency tone for $T_{\text{capture}}$ seconds each. $T_{\text{capture}}$ is chosen to be long enough to contain at least two repetitions of the target radar's frame, whose interval $T_{\text{frame}}$ is unknown to the spoofer but can be reasonably bounded. The samples from both captures are passed through the same CFAR detector, creating two sets of detections: $\mathcal{D}_1$ and $\mathcal{D}_2$. Note that $f_1=f_2=f_{\text{tone}}$ since the mixing frequency has not changed. These detections are visualized in relation to the frame parameters in Fig.~\ref{fig:params_frame} and to the chirp parameters in Fig.~\ref{fig:params_chirp}. Since the spoofer currently has no reference of the target's frame timing, the spoofer does not yet know which frame each detection belongs to. To estimate the frame timing under this ambiguity, the spoofer analyzes the periodicity of the detections. This is done by evaluating how well these two sets match over shifts in time using a Gaussian kernel likelihood function $k(t_1,t_2,\sigma) = \exp{\left(\frac{-|t_1 - t_2|^2}{\sigma^2}\right)}$. This approach is inspired by kernel correlation techniques used in point-set registration \cite{tsin2004correlation}. The sum kernel likelihood between the sets for a lag $l$ is
	\begin{equation}
		d(l) = \sum_{(t_1,f_1) \in \mathcal{D}_1} \sum_{(t_2,f_2) \in \mathcal{D}_2} k(t_1,t_2-l,\sigma).
	\end{equation}
	
\begin{figure}
		\centering
		\begin{subfigure}{.47\linewidth}
			\centering
			\includegraphics[width=0.99\linewidth,page=1,trim={0 0 0 0},clip]{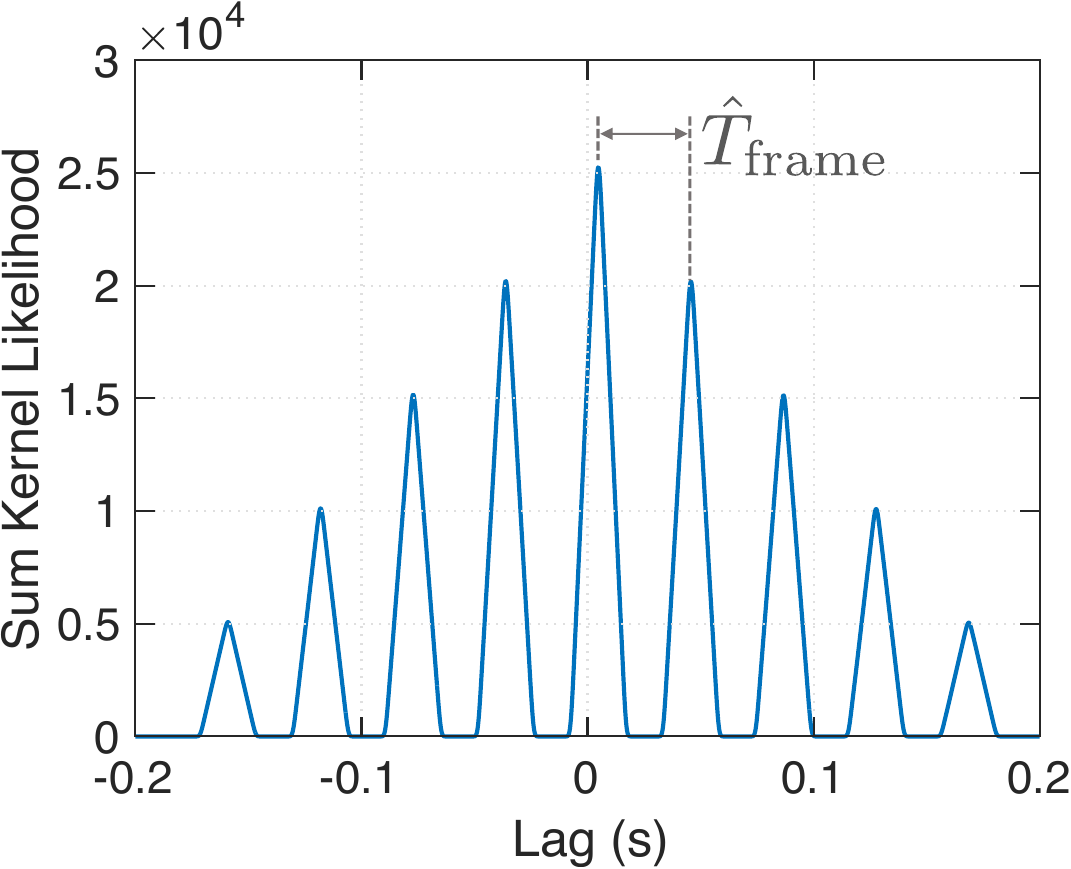}
			\caption{}
			\label{fig:frame_timing}
		\end{subfigure}%
		\begin{subfigure}{.53\linewidth}
			\centering
			\includegraphics[width=0.99\linewidth,page=1]{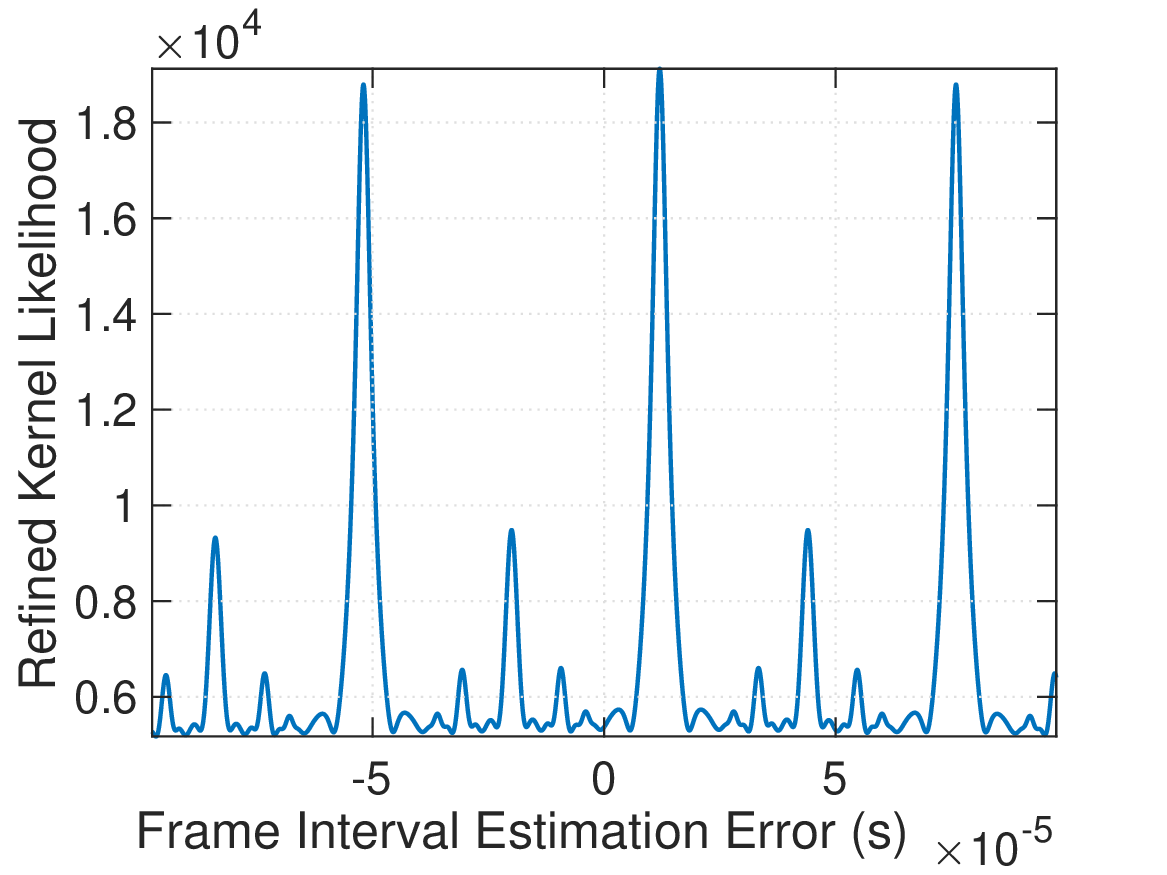}
			\caption{}
			\label{fig:refined_kernel_dist}
		\end{subfigure}
		\caption{(a) The sum kernel likelihood between $\mathcal{D}_1$ and $\mathcal{D}_2$ for different lag times. The distance between each peak provides an estimate of the frame interval. (b) The refined kernel likelihood used for estimating the error in the initial frame interval estimate.}
	\end{figure}

	The kernel standard deviation $\sigma$ is chosen to be greater than the longest possible chirp interval and less than the shortest possible frame interval, causing the function $d(l)$ to smooth over small time differences between the two sets. The result is a function that exhibits multiple peaks with an inter-peak distance approximately equal to the frame interval, from which one may obtain an estimate $\hat{T}_{\text{frame}}$. Fig.~\ref{fig:frame_timing} shows an example of this sum kernel likelihood function.
	
	The spoofer then estimates the frame timing offset $\Delta t_{\text{frame}}$. Since there is an inherent ambiguity between the frame timing offset and the chirp timing offsets relative to the frame start, the spoofer has flexibility in selecting this value. However, it is important that the chosen offset does not cause the estimated frames to contain detections from two of the target's frames. To minimize this possibility, $\Delta t_{\text{frame}}$ is chosen as the time that minimizes the mean squared distance between the detection times and the center of the nearest frame. The two sets of detections are combined to form $\mathcal{D} = \mathcal{D}_1 \cup \mathcal{D}_2$, and the frame offset $\Delta t_{\text{frame}}$ is estimated as
	\begin{align}
		&\Delta \hat{t}_{\text{frame}} = \\
		&\operatorname*{argmin}_{\Delta t_{\text{frame}} \in [0,\hat{T}_{\text{frame}}]} \sum_{(t,f) \in \mathcal{D}} \left|((t - \Delta t_{\text{frame}}) \Mod{\hat{T}_{\text{frame}}}) - \frac{\hat{T}_{\text{frame}}}{2} \right|^2. \nonumber
	\end{align}
	
	With an estimate of the frame interval and a frame time offset that centers the chirp sequence in the frame, the spoofer can now partition its detections into each frame and further refine its frame interval estimate. Similar to the first frame interval estimation, a kernel likelihood measure is used. First, define a function $g(t,l) = (t - \Delta \hat{t}_{\text{frame}}) \Mod (\hat{T}_{\text{frame}}-l)$, and let $\sigma_{\text{ref}}$ be a refined kernel standard deviation, which is chosen to be less than the minimum chirp duration but larger than the spoofer's desired time resolution. Then the sum kernel likelihood using the refined kernel is
	\begin{equation}
		d_{\text{ref}}(l) = \sum_{(t_1,f_1) \in \mathcal{D}} \sum_{(t_2,f_2) \in \mathcal{D}} k(g(t_1,l), g(t_2,l), \sigma_{\text{ref}}).
	\end{equation}
	
	The frame interval estimate error is given by the value $l$ that maximizes the refined kernel likelihood: $\hat{l}_{\text{max}} = \operatorname*{argmax}_{l} d_{\text{ref}}(l)$. This likelihood function is visualized in Fig.~\ref{fig:refined_kernel_dist}. Using this error, the spoofer creates a new set of detections that accounts for the refined frame timing estimate:
	\begin{align}
		\mathcal{D}_{\text{frame}} = \{(g(t,\hat{l}_{\text{max}}),f)\ :\ (t,f)\in\mathcal{D}\}.
	\end{align}
	Finally, the frame interval estimate $\hat{T}_{\text{frame}}$ is adjusted by $\hat{l}_{\text{max}}$ seconds for future captures.  Frame timing estimation is complete now that the spoofer has estimates $\hat{T}_{\text{frame}}$ and $\Delta \hat{t}_{\text{frame}}$.
	
	Next, the spoofer identifies each chirp in the frame. Chirps from different frames are matched by their proximity in time in the new set $\mathcal{D}_{\text{frame}}$. The matching criterion for $(t_1,f_1) \in \mathcal{D}_{\text{frame}}$ and $(t_2,f_2) \in \mathcal{D}_{\text{frame}}$ is $\left(t_1-t_2\right)^2 < \sigma_{\text{chirp}}^2$. Each unique set of matching detection times is considered a detected chirp, creating an initial estimate of the set of chirps $\hat{K}$. Each chirp's time offset is set equal to the average of the detection times matched to that chirp. The chirp timing threshold $\sigma_{\text{chirp}}$ is chosen to be larger than the worst-case timing jitter and less than the minimum inter-chirp time.
	
	\begin{figure}
		\centering
		\includegraphics[width=0.7\linewidth,page=9]{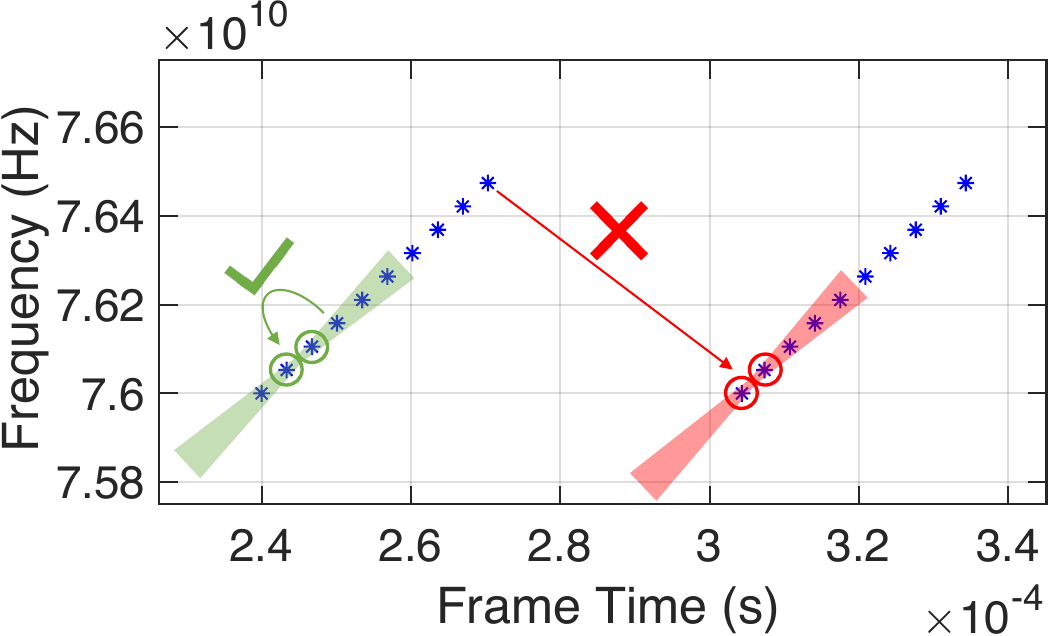}
		\caption{A visualization of the detections obtained after a frequency search and how the chirp estimation algorithm operates. The circles indicate detections already associated with a chirp, the cones indicate the acceptable likelihood regions for associating new detections, and the arrows indicate prospective detections to be associated with each chirp. On the leftmost chirp, the prospective detection is accepted. On the rightmost chirp, the prospective detection is excluded.}
		\label{fig:chirp_est_alg}
	\end{figure}
	
	
	Now that the spoofer has a sense of the target radar's frame timing and chirp locations, the spoofer can begin sampling frequencies to determine each chirp's duration $T_k$, start time $\Delta t_k$, start frequency $\Delta f_k$, and slope $\beta_k$. The spoofer captures samples for $M_{\text{search}}$ frames, mixing with a different constant frequency tone $f_m,\ \forall m \in \{1, \ldots, M_{\text{search}}\}$ in the $m$th frame. In this paper, $f_m$ were linearly spaced in frequency, but other strategies are possible. As before, the spoofer creates a set of detections $D_{\text{search}}$, from which it estimates each chirp's parameters using Algorithm \ref{alg:chirp_est}, which associates likely detections with each chirp, performs a least-squares regression, and updates the covariance of its parameter estimates. This process is visualized in Fig.~\ref{fig:chirp_est_alg}. The inputs to the algorithm are an initial estimated time for the chirp $\Delta t_{k}$, the set of detections $D_{\text{search}}$, and a prior measurement variance for detection frequencies $\sigma_{\text{freq}}^2$. The output is a chirp slope estimate $\hat{\beta}_k$, a linear regression frequency offset $\hat{f}_k$, a vector of detection times $\bm{t}_{\text{match},k}$ associated with the chirp, and a vector of frequencies $\bm{f}_{\text{match},k}$ associated with the chirp. This algorithm is run for each chirp $k \in \hat{K}$. Let $n_{\text{start},k} = \operatorname*{argmin}_n \left(\bm{t}_{\text{match},k}\right)_n$, the index of the first occuring matched detection. Then the $k$th chirp's start frequency is estimated as $\Delta \hat{f}_k = \left(\bm{f}_{\text{match},k}\right)_n$. The start time is estimated using the linear regression frequency offset as $\Delta \hat{t}_k = (\Delta \hat{f}_k - \hat{f}_k)/\beta_k$. The spoofer now has an estimate of all signal parameters required to recreate the target signal accurately enough to begin tracking.
	
	\begin{algorithm}
		\caption{Iterative Chirp Parameter Estimation}\label{alg:chirp_est}
		\small
		\hspace*{\algorithmicindent}\textbf{Input}: $\Delta{t}_{k},\ D_{\text{search}},\ \sigma_{\text{freq}}^2$\\
		\hspace*{\algorithmicindent}\textbf{Output}: $\hat{\beta}_k,\ \hat{f}_k,\ \bm{t}_{\text{match},k},\ \bm{f}_{\text{match},k}$
		
		\begin{algorithmic}[1]
			\STATE $\hat{\beta}_k \leftarrow 0,\ \hat{f}_k \leftarrow 0,\ \bm{K} \leftarrow \text{diag}\{[\infty, \infty]\}$
			\STATE $\mathcal{D}^\prime\leftarrow$ $D_{\text{search}}$ sorted by distance from $\Delta{t}_{k}$
			\STATE $(t_1,f_1) \leftarrow \text{1st entry in } \mathcal{D}^\prime,\ \mathcal{D}^\prime \leftarrow \mathcal{D}^\prime\setminus (t_1,f_1)$
			\STATE $\bm{t}_{\text{match}} \leftarrow [t_1],\ \bm{f}_{\text{match}} \leftarrow [f_1]$
			\FOR{$\{t,f\} \in \mathcal{D}^\prime$}
			\STATE $\hat{f}_k \leftarrow \hat{\beta}_k t + f$
			\STATE $\bm{x} \leftarrow [t\ 1]$
			\STATE $\hat{\sigma}_{\text{freq}}^2 \leftarrow \bm{x}\bm{K}\bm{x}^H + \sigma_{\text{freq}}^2$
			\IF{$\frac{|f-\hat{f}_k|}{\hat{\sigma}_{\text{freq}}} < 2$}
			\STATE $\bm{t}_{\text{match},k} \leftarrow [\bm{t}_{\text{match},k}^{\top}\ t]^{\top},\ \bm{f}_{\text{match},k} \leftarrow [\bm{f}_{\text{match},k}^{\top}\ f]^{\top}$
			\STATE $\bm{X} \leftarrow [\bm{t}_{\text{match},k}\ \bm{1}]$
			\STATE $[\hat{\beta}_k\ \hat{f}_k]^{\top} \leftarrow (\bm{X}^H\bm{X})^{-1}\bm{X}^H \bm{f}_{\text{match},k}$
			\STATE $\bm{K} \leftarrow \sigma_{\text{freq}}^2 (\bm{X}^H\bm{X})^{-1}$
			\ELSE
			\STATE exit
			\ENDIF
			\ENDFOR
		\end{algorithmic}
	\end{algorithm}
	
	\subsection{Tracking}
	
	After identifying the target radar's signal as described, the spoofer can track the signal using more traditional FMCW signal processing, similar to the processing described in \cite{patole2017automotive}. During this tracking phase, the individual chirp parameters are refined to produce a mixed signal $m_{l\tilde{k}}^{ikr}(t)$ similar to what would arise when tracking a genuine reflection. Rather than using continuous frequency tones, as in the identification stage, the spoofer uses its estimated target signal in its mixer.
	
	Recall that the spoofer collects a vector of samples during its frame denoted as $\bm{y}_{i}[n]$. Rather than sampling continously over the entire frame, the spoofer now only samples during each chirp. The vector of sample indices corresponding to chirp $k$ can be denoted as $\bm{n}_{k}$, containing $N_{\text{sample}}$ samples, assuming the same number of samples are collected during each chirp. These samples can be structured into a 3-dimensional tensor $(\bm{Y}_{i}^{(k)})_{an} = (\bm{y}_{i}[(\bm{n}_{k})_n])_a$, where the 1st dimension is the receive antenna index $a$, the 2nd dimension is the sample index $n$, and the 3rd dimension (denoted with the superscript) is the chirp index $k$. In the radar literature, this tensor is commonly referred to as the ``radar data cube.''
	
	The spoofer beamforms across receive antennas with an FFT across the 1st dimension of the radar data cube, creating $\tilde{\bm{Y}}_{i}^{(k)}$. The spoofer also performs range and Doppler processing with an FFT, as in \cite {patole2017automotive}, over the 2nd and 3rd dimensions of this tensor, creating $\breve{\bm{Y}}_{i}^{(d)}$, where $d$ is the post-FFT Doppler bin index. Note that this FFT processing assumes homogenous chirp slopes. For non-homogeneous chirps, the FFTs must be transformed to align in range, the details of which are omitted in this paper. Next, the spoofer determines the beamforming, range, and Doppler bin indices ($b$, $n$, and $d$) that maximize the power of this processed tensor:
	\begin{align}
		\hat{b}_{\text{max}},\hat{n}_{\text{max}},\hat{d}_{\text{max}} = \operatorname*{argmax}_{b,n,d} \left|\left(\breve{\bm{Y}}_{i}^{(d)}\right)_{bn}\right|^2.
		\label{eq:peak}
	\end{align}
	These indices are then converted into an angle $\hat{\theta}$ in radians, a time delay $\Delta \hat{t}$ in seconds, and a Doppler frequency $\hat{f}_{\text{D}}$ in \SI{}{\hertz}. The spoofer implements a simple 1st-order tracker to place the peak at a desired time delay and Doppler frequency.
	
	Due to inaccurate chirp timing and chirp slope estimates, the processed signal may appear spread out in range and Doppler, reducing or eliminating any clear peak. To address this, the spoofer first refines its individual chirp estimates using the beamformed signal prior to range and Doppler processing $\tilde{\bm{Y}}_{i}^{(k)}$. The spoofer takes the $\hat{b}_{\text{max}}$th row of $\tilde{\bm{Y}}_{i}^{(k)}$, denoted $(\bm{z}_{i}^{(k)})^{\top}$, to beamform in the target's direction. Then, it performs only range processing through an FFT of this vector, creating $\breve{\bm{z}}_{i}^{(k)}$. This differs from $\breve{\bm{Y}}_{i}^{(d)}$ since only the power-maximizing beam is used and Doppler processing is not performed, creating a separate range signal for each chirp. The range index $n$ with the peak power is found as
	\begin{equation}
		\hat{n}_{\text{max},k} = \operatorname*{argmax}_{n} \left|\left( \breve{\bm{z}}_{i}^{(k)} \right)_{n}\right|^2.
	\end{equation}
	
	The complex values at these peak-power range indices are $p_k = ( \breve{\bm{z}}_{i}^{(k)} )_{\hat{n}_{\text{max},k}}$. Then, the range indices $\hat{n}_{\text{max},k}$ are converted into time delay estimates $\delta \hat{t}_k$. Let $\delta \bar{t}$ be the average of $\delta \hat{t}_k$ over $k$. The spoofer shifts the start time of the $k$th chirp by $\delta \hat{t}_k - \delta \bar{t}$ seconds to align all chirps at the same range.
	
	\begin{figure}
		\centering
		\begin{subfigure}{0.48\linewidth}
			\includegraphics[width=0.99\linewidth,trim={0.2cm 0 0.5cm 0},clip]{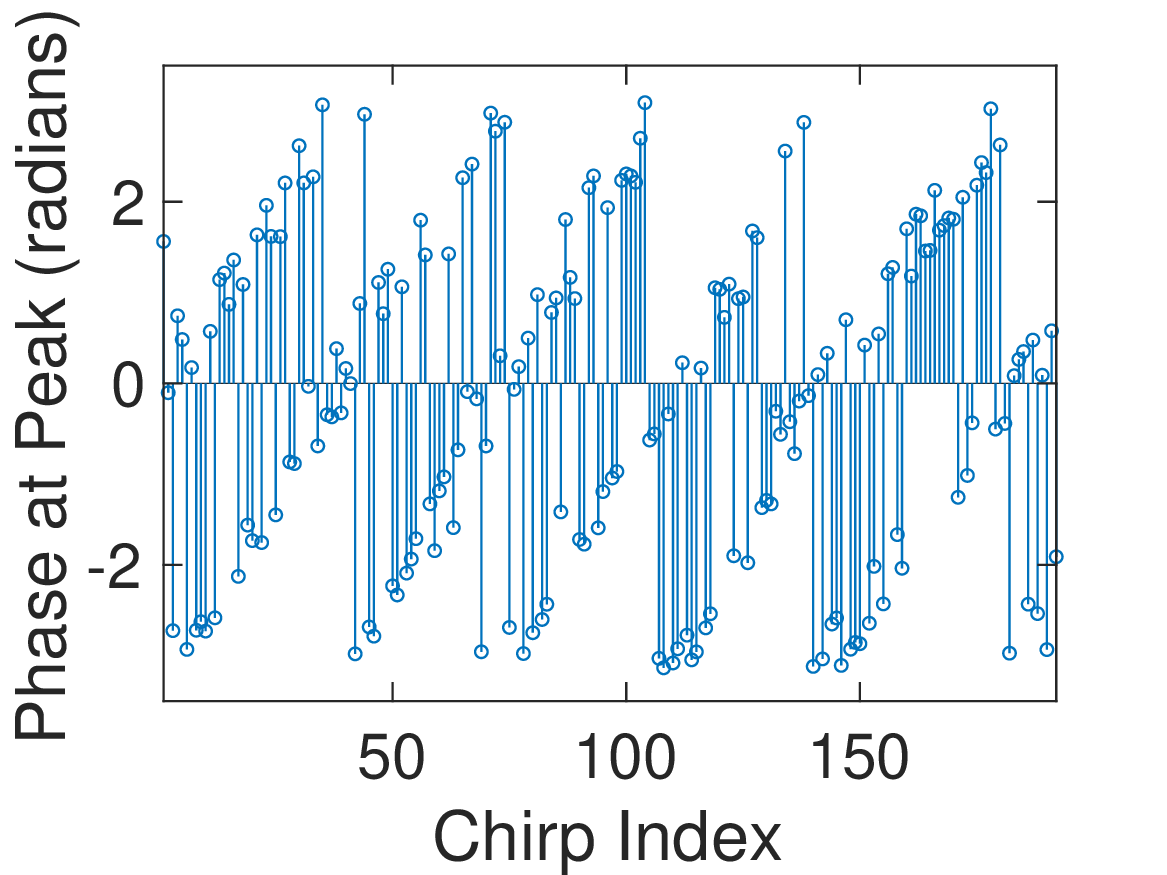}
			\caption{}
			\label{fig:chirp_phases}
		\end{subfigure}
		\begin{subfigure}{0.48\linewidth}
			\includegraphics[width=0.99\linewidth,trim={0.2cm 0 0.5cm 0},clip]{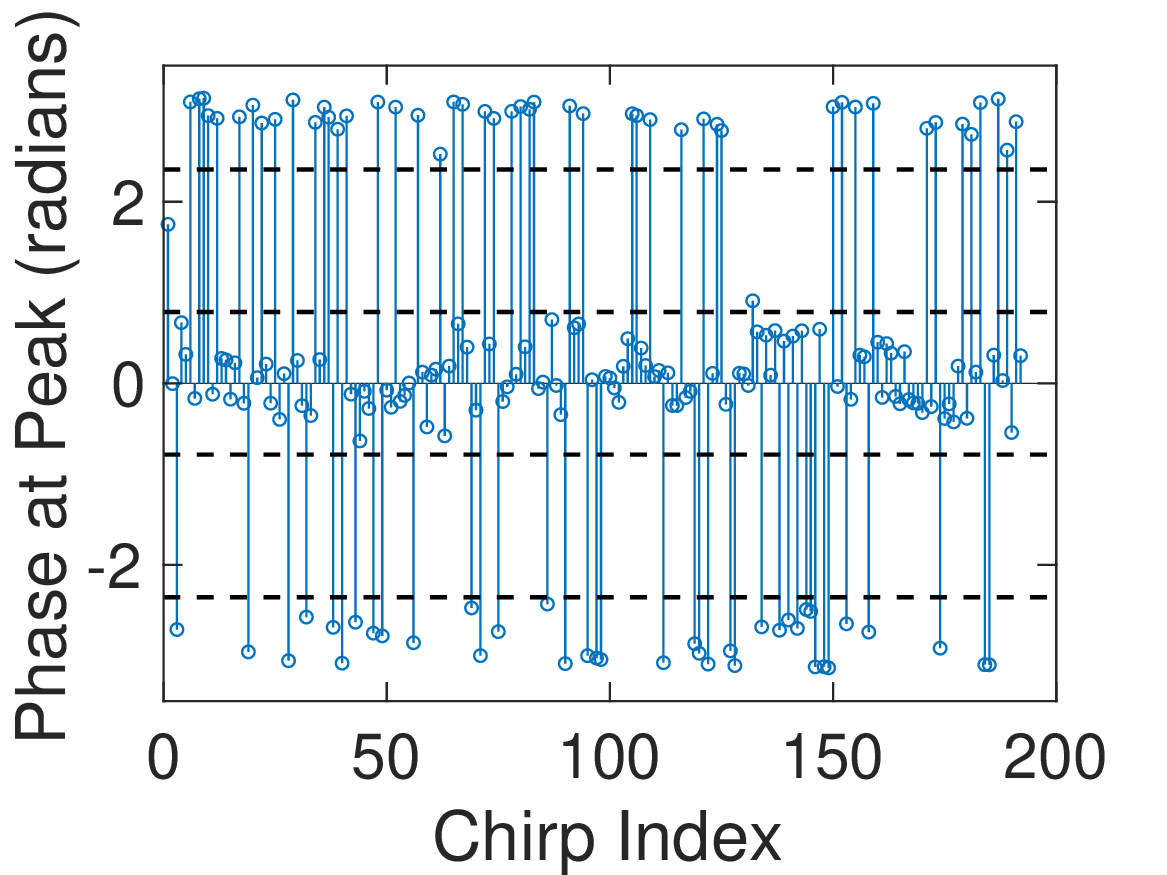}
			\caption{}
			\label{fig:chirp_phases_vv}
		\end{subfigure}
		\caption{The estimated chirp phases at the peak power bin after range processing and beamforming, prior to (a) and after (b) correction. The phases exhibit a phase rotation over time due to residual Doppler while also showing a random \SI{180}{\degree} phase shift due to the coding. After correction, the phases can be quantized based on the detection regions shown.}
	\end{figure}
	
	Additionally, the spoofer can refine its estimate of each chirp's slope. By (\ref{eq:mix_sig_passive}), slope error causes the mixed signal to have a residual linear frequency slope. This is addressed by searching for the chirp slope rate that, when mixed with the samples prior to range processing, maximizes the power in the peak of the range FFT. For a residual slope $\hat{\beta}$, the slope-correction signal is $m_{\hat{\beta}}[n] = \exp ( j\pi \hat{\beta}(T_s n)^2 ),\ \forall n \in \{1, \ldots, N_{\text{sample}}\}$, denoted in vector notation as $\bm{m}_{\hat{\beta}}$. This is mixed with the beamformed sample signal creating $\bm{z}_{i\hat{\beta}}^{(k)}  = \bm{z}_{i}^{(k)} \odot \bm{m}_{\hat{\beta}}$ (with $\odot$ being a Hadamard product). Then, the spoofer performs its range FFT, creating $\breve{\bm{z}}_{i\hat{\beta}}^{(k)}$. The residual slope is estimated as
	\begin{align}
		\delta \hat{\beta}_k = \operatorname*{argmax}_{\beta} \max_{n} \left|\left(\breve{\bm{z}}_{i\beta}^{(k)}\right)_{n}\right|^2.
	\end{align}
	
	The spoofer then adjusts the slope for the $k$th chirp by $-\delta\hat{\beta}_k$. The last chirp-level parameters to be estimated and compensated for are the chirp phase shifts $\phi_k,\ \forall k \in \hat{K}$, which arise from both transmit phase weightings and the physical placement of transmit antennas. Recall the values $p_k$ obtained earlier, containing the complex values at the power-maximizing range bin for each chirp, which can be stacked into a vector $\bm{p}$. Fig.~\ref{fig:chirp_phases} shows an example of the phases of $\bm{p}$ where residual Doppler is present, despite the Doppler tracking. Therefore, a Viterbi and Viterbi (V\&V) Algorithm \cite{viterbi1983nonlinear} is used to wipe off any phase modulation for refined Doppler estimation. Assuming a modulation order of $M_{\text{mod}}$ which is a power of 2, the modified values are $p_{\text{V},k} = p_{k}^{\log_2(M_{\text{mod}})}$. The spoofer then finds the index corresponding to the peak of the FFT of $p_{\text{V},k}$. This index is converted into a frequency estimate $\hat{f}_{\text{V}}$ and a phase estimate $\hat \phi_{\text{V}}$. Let $T_c$ be the average chirp interval. A correction signal is created $\hat{m}_{\text{V}}[n] = \exp ( j2\pi \hat{f}_{\text{V}} T_c k + j \hat \phi_{\text{V}}),\ \forall k \in \hat{K}$, denoted in vector notation as $\hat{\bm{m}}_{\text{V}}$. This signal is multiplied with $\bm{p}$, giving $\bm{p}_{\text{corr}} = \bm{p} \odot \hat{\bm{m}}_{\text{V}}$. In this new signal, phase shifts due to any residual Doppler are approximately removed, and the relative phase shift of each chirp can be estimated. The spoofer decodes the phase modulation of each chirp through detection regions defined by the modulation order $M_{\text{mod}}$, creating an estimate $\hat{\phi}_k$. This FFT-based approach relies on the chirp spacing being approximately uniform, but could be extended to use a non-uniform DFT when chirp spacing is non-uniform. Fig.~\ref{fig:chirp_phases_vv} shows the phases of $\bm{p}_{\text{corr}}$ derived from the same $\bm{p}$ shown in Fig.~\ref{fig:chirp_phases} with QPSK detection regions overlaid.
	
	\section{Results}
	\label{sec:results}
	
	\begin{figure}
		\centering
		\begin{subfigure}{1\linewidth}
			\centering
			\includegraphics[width=0.82\linewidth,trim={0 2.3cm 0cm 3.7cm},clip]{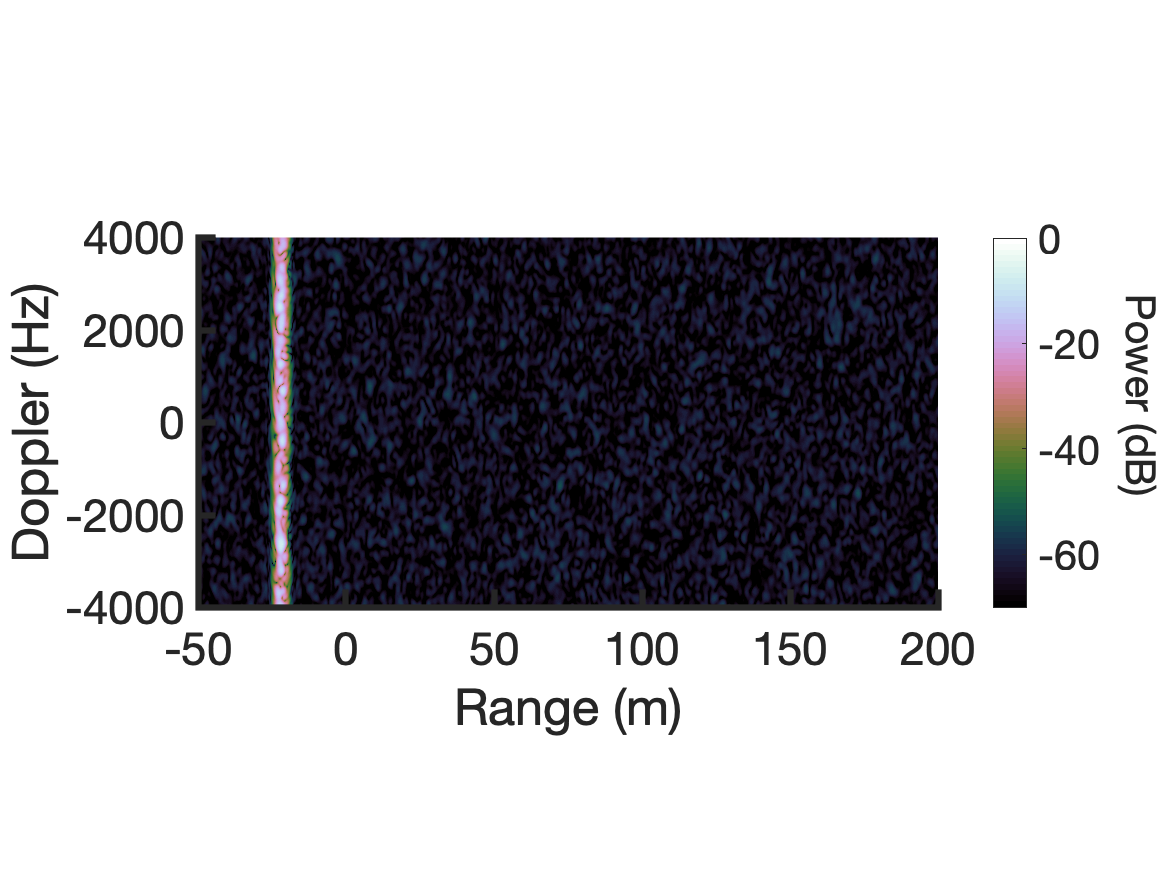}
			\caption{}
			\label{fig:rd41}
		\end{subfigure}\\
		\begin{subfigure}{1\linewidth}
			\centering
			\includegraphics[width=0.82\linewidth,trim={0cm 2.3cm 0cm 3.7cm},clip]{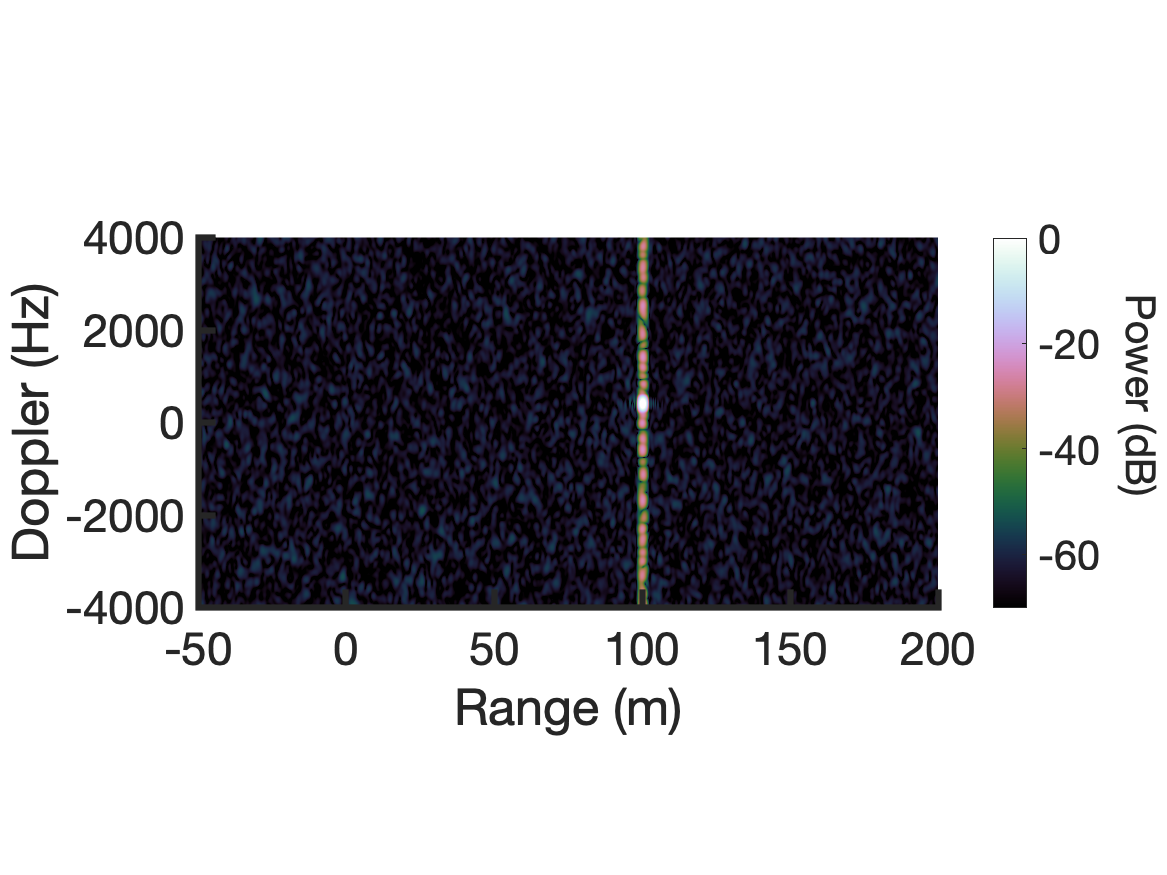}
			\caption{}
			\label{fig:rd52}
		\end{subfigure}
		\caption{The spoofer's range-Doppler processing in the 1st (a) and 12th (b) frames of the tracking stage. The signal is highly spread in range and Doppler in the 1st frame. The Doppler spreading is refined by correcting the chirp slope, timing, and phase code in the 12th frame, producing a clear peak.}
	\end{figure}
	
	The results shown are from MATLAB simulation. The simulated scenario has one target radar, one spoofing radar, and one point reflector colocated with the target radar. The target radar transmits a frame with a standard time-multiplexed MIMO configuration across 3 transmit and 4 receive antennas. The IF bandwidth is \SI{10}{\mega\hertz} and the target's chirps sweep a bandwidth of \SI{500}{\mega\hertz} in \SI{32}{\micro\second}. A random binary phase code is applied to each chirp's weighting which repeats each frame. The spoofer radar is equipped with the same antenna elements and RF chains, but has 8 receive antennas. The radars are placed \SI{50}{\meter} apart and have a relative velocity of \SI{14.14}{\meter\per\second}.
	
\begin{figure}
		\centering
		\begin{subfigure}{.47\linewidth}
			\centering
			\includegraphics[width=1.0\linewidth,trim={0.2cm 0 0cm 0.4cm},clip]{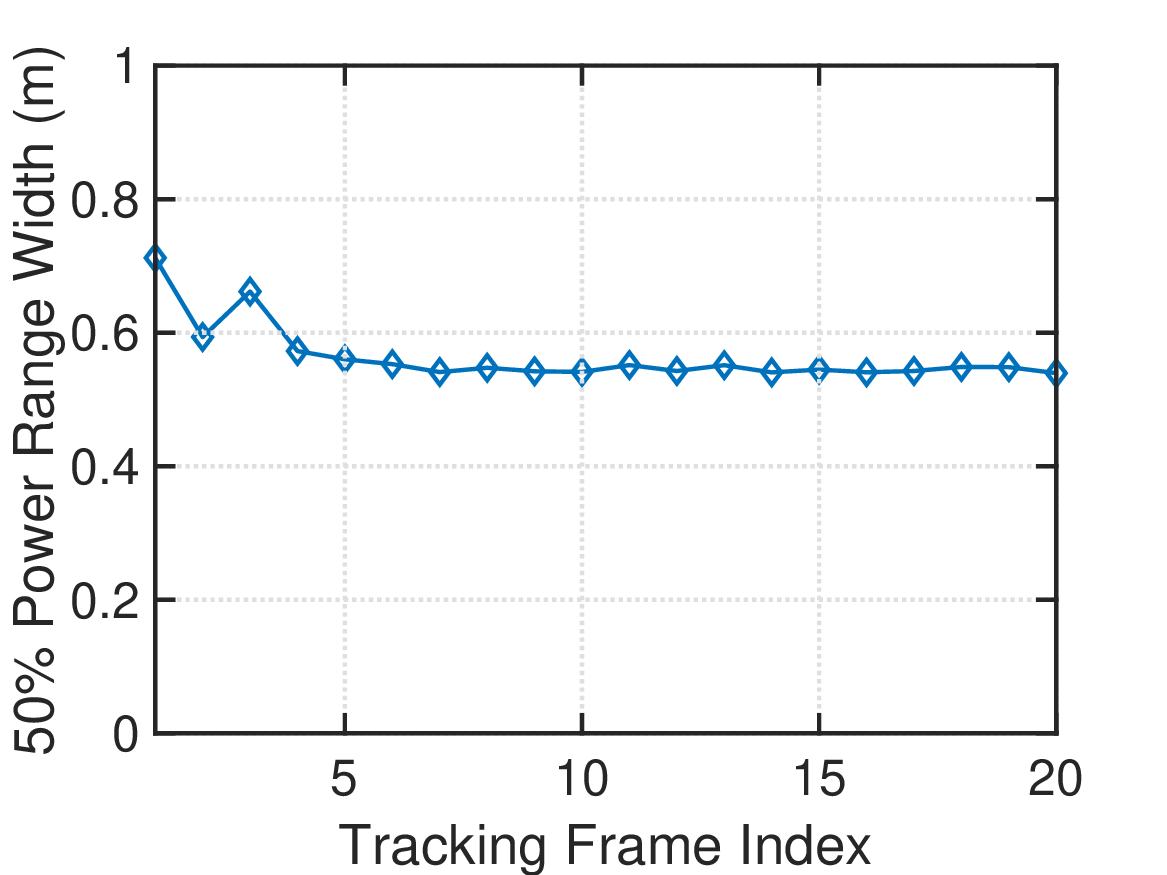}
			\caption{}
			\label{fig:range_width}
		\end{subfigure}%
		\begin{subfigure}{.47\linewidth}
			\centering
			\includegraphics[width=1.0\linewidth,trim={0cm 0 0.2cm 0.4cm},clip]{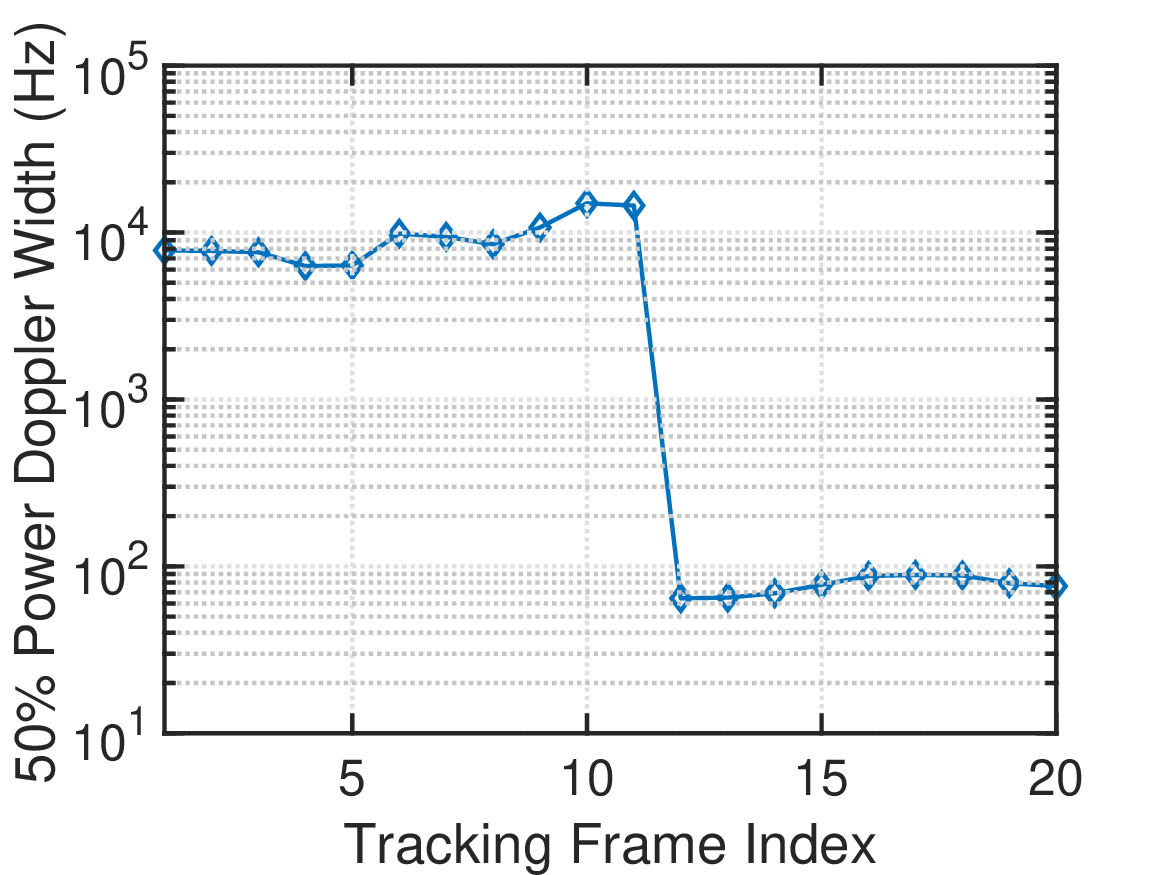}
			\caption{}
			\label{fig:doppler_width}
		\end{subfigure}\\
		\begin{subfigure}{0.47\linewidth}
			\centering
			\includegraphics[width=1.0\linewidth,trim={0.1cm 0 0.1cm 0.6cm},clip]{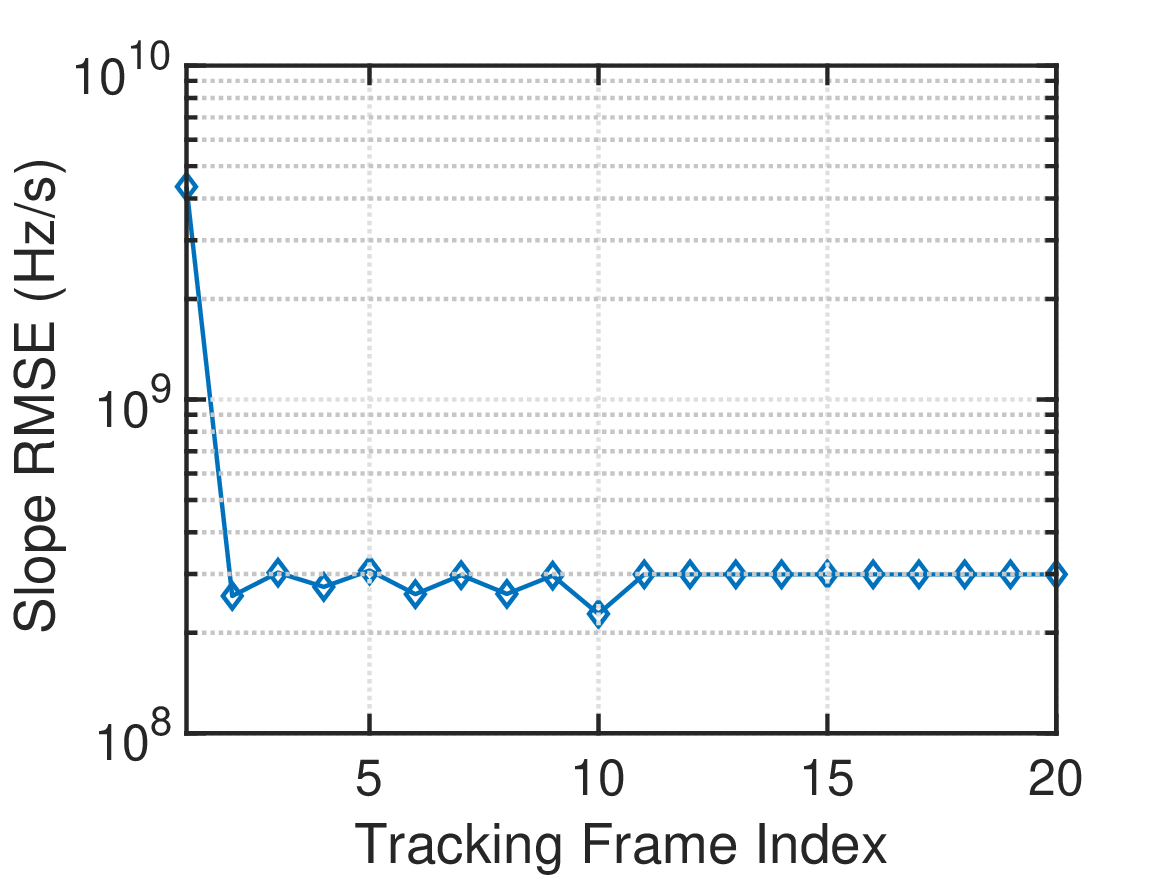}
			\caption{}
			\label{fig:slope_rmse}
		\end{subfigure}
		\caption{(a) The width around the peak of the range-Doppler map containing \SI{50}{\percent} of the signal power in the range dimension. (b) The width around the peak of the range-Doppler map containing \SI{50}{\percent} of the signal power in the Doppler dimension. (c) The RMSE of the chirp slope estimates.}
	\end{figure}

	Fig.~\ref{fig:rd41} and Fig.~\ref{fig:rd52} show the range-Doppler map that the spoofer perceives during the tracking stage. In Fig.~\ref{fig:rd41}, the spoofer has just finished its identification stage. Due to coarse estimates of the chirp timings and slopes, the power is spread out in range and Doppler. Furthermore, the spoofer's tracker aims to place this signal at a perceived range of \SI{100}{\meter} and Doppler of \SI{0}{\hertz}. In Fig.~\ref{fig:rd52}, the tracker has shifted the signal to its desired range and refined its individual chirp timing and slope estimates, concentrating the power in range. The tracker has also corrected for its chirp phase estimates, concentrating the power in Doppler. The performance of the tracker can be quantified with three metrics: power concentration in range, power concentration in Doppler, and RMSE of the chirp slope estimates. Fig.~\ref{fig:range_width} shows the concentration in range, measured by the width around the peak that contains \SI{50}{\percent} of the signal's power. As the chirp timing and slope are refined, this width reduces to near \SI{0.54}{\meter}. Fig.~\ref{fig:doppler_width} shows the same analysis for Doppler. No improvement in Doppler spread is seen until frame $12$ when the phase estimates are applied, decreasing the spread to near \SI{63}{\hertz}. After frame $12$, some fluctuation is seen due to TX-dependent phase shifts from the array manifold as the target radar moves.
	
	
	\section{Conclusion}
	\label{sec:conclusion}
	
	These results demonstrate how an intelligent adversary may succesfully identify and track a mmWave FMCW radar using a COTS FMCW radar itself. This adversary does not require any prior knowledge of the target's signal's timing and chirp parameters, making such an attack effective in the real-world where the target's signal is unknown. After the described entrainment, the spoofer may transmit its signal, inducing the target to perceive very precisely controlled spoofed objects. In future analysis, it will be crucial to consider such an adversary when evaluating spoofing countermeasures.
	
	\section*{Acknowledgment}
	
	Research was sponsored by the U.S. Department of Transportation (USDOT) under the University Transportation Center (UTC) Program Grant 69A3552047138 (CARMEN), and by affiliates of the 6G@UT center within the Wireless Networking and Communications Group at The University of Texas at Austin.
	
	\bibliographystyle{IEEEtran}
	{
		\bibliography{pangea}

\begin{thebibliography}{10}
\providecommand{\url}[1]{#1}
\csname url@samestyle\endcsname
\providecommand{\newblock}{\relax}
\providecommand{\bibinfo}[2]{#2}
\providecommand{\BIBentrySTDinterwordspacing}{\spaceskip=0pt\relax}
\providecommand{\BIBentryALTinterwordstretchfactor}{4}
\providecommand{\BIBentryALTinterwordspacing}{\spaceskip=\fontdimen2\font plus
\BIBentryALTinterwordstretchfactor\fontdimen3\font minus
  \fontdimen4\font\relax}
\providecommand{\BIBforeignlanguage}[2]{{%
\expandafter\ifx\csname l@#1\endcsname\relax
\typeout{** WARNING: IEEEtran.bst: No hyphenation pattern has been}%
\typeout{** loaded for the language `#1'. Using the pattern for}%
\typeout{** the default language instead.}%
\else
\language=\csname l@#1\endcsname
\fi
#2}}
\providecommand{\BIBdecl}{\relax}
\BIBdecl

\bibitem{hasch2012millimeter}
J.~Hasch, E.~Topak, R.~Schnabel, T.~Zwick, R.~Weigel, and C.~Waldschmidt,
  ``Millimeter-wave technology for automotive radar sensors in the 77 {GHz}
  frequency band,'' \emph{IEEE Transactions on Microwave Theory and
  Techniques}, vol.~60, no.~3, pp. 845--860, 2012.

\bibitem{sun2020mimo}
S.~Sun, A.~P. Petropulu, and H.~V. Poor, ``{MIMO} radar for advanced
  driver-assistance systems and autonomous driving: {Advantages} and
  challenges,'' \emph{IEEE Signal Processing Magazine}, vol.~37, no.~4, pp.
  98--117, 2020.

\bibitem{patole2017automotive}
S.~M. Patole, M.~Torlak, D.~Wang, and M.~Ali, ``Automotive radars: A review of
  signal processing techniques,'' \emph{IEEE Signal Processing Magazine},
  vol.~34, no.~2, pp. 22--35, 2017.

\bibitem{engels2017advances}
F.~Engels, P.~Heidenreich, A.~M. Zoubir, F.~K. Jondral, and M.~Wintermantel,
  ``Advances in automotive radar: A framework on computationally efficient
  high-resolution frequency estimation,'' \emph{IEEE Signal Processing
  Magazine}, vol.~34, no.~2, pp. 36--46, 2017.

\bibitem{lies2021longRange}
W.~A. Lies, L.~Narula, P.~A. Iannucci, and T.~E. Humphreys, ``Long-range, low
  {SWaP-C} {FMCW} radar,'' \emph{IEEE Journal of Selected Topics in Signal
  Processing}, pp. 1--1, 2021.

\bibitem{D12mosarim2010}
M.~Ahrholdt \emph{et~al.}, ``D12.1-study report on relevant scenarios and
  applications and requirements specification,'' \emph{European Commission:
  MOre Safety for All by Radar Interference Mitigation (MOSARIM)}, 2010.

\bibitem{D1mosarim2010}
M.~Kunert \emph{et~al.}, ``D1.5-study on the state-of-the-art interference
  mitigation techniques,'' \emph{European Commission: MOre Safety for All by
  Radar Interference Mitigation (MOSARIM)}, 2010.

\bibitem{alland2019interference}
S.~Alland, W.~Stark, M.~Ali, and M.~Hegde, ``Interference in automotive radar
  systems: Characteristics, mitigation techniques, and current and future
  research,'' \emph{IEEE Signal Processing Magazine}, vol.~36, no.~5, pp.
  45--59, 2019.

\bibitem{komissarov2021spoofing}
R.~Komissarov and A.~Wool, ``Spoofing attacks against vehicular {FMCW} radar,''
  in \emph{Proceedings of the 5th Workshop on Attacks and Solutions in Hardware
  Security}, 2021, pp. 91--97.

\bibitem{miura2019low}
N.~Miura, T.~Machida, K.~Matsuda, M.~Nagata, S.~Nashimoto, and D.~Suzuki, ``A
  low-cost replica-based distance-spoofing attack on {mmWave} {FMCW} radar,''
  in \emph{Proceedings of the 3rd ACM Workshop on Attacks and Solutions in
  Hardware Security Workshop}, 2019, pp. 95--100.

\bibitem{ordean2022millimeter}
M.~Ordean and F.~D. Garcia, ``Millimeter-wave automotive radar spoofing,''
  \emph{arXiv preprint arXiv:2205.06567}, 2022.

\bibitem{nallabolu2021frequency}
P.~Nallabolu and C.~Li, ``A frequency-domain spoofing attack on {FMCW} radars
  and its mitigation technique based on a hybrid-chirp waveform,'' \emph{IEEE
  Transactions on Microwave Theory and Techniques}, vol.~69, no.~11, pp.
  5086--5098, 2021.

\bibitem{sun2021control}
Z.~Sun, S.~Balakrishnan, L.~Su, A.~Bhuyan, P.~Wang, and C.~Qiao, ``Who is in
  control? {P}ractical physical layer attack and defense for {mmWave}-based
  sensing in autonomous vehicles,'' \emph{IEEE Transactions on Information
  Forensics and Security}, vol.~16, pp. 3199--3214, 2021.

\bibitem{gardill2019situ}
M.~Gardill, J.~Schwendner, and J.~Fuchs, ``In-situ time-frequency analysis of
  the 77 {GHz} bands using a commercial chirp-sequence automotive {FMCW} radar
  sensor,'' in \emph{2019 IEEE MTT-S International Microwave Symposium
  (IMS)}.\hskip 1em plus 0.5em minus 0.4em\relax IEEE, 2019, pp. 544--547.

\bibitem{gardill2020approach}
------, ``An approach to over-the-air synchronization of commercial
  chirp-sequence automotive radar sensors,'' in \emph{2020 IEEE Topical
  Conference on Wireless Sensors and Sensor Networks (WiSNeT)}.\hskip 1em plus
  0.5em minus 0.4em\relax IEEE, 2020, pp. 46--49.

\bibitem{roehr2007radiomethod}
S.~Roehr, P.~Gulden, and M.~Vossiek, ``Method for high precision clock
  synchronization in wireless systems with application to radio navigation,''
  in \emph{2007 IEEE Radio and Wireless Symposium}.\hskip 1em plus 0.5em minus
  0.4em\relax IEEE, 2007, pp. 551--554.

\bibitem{lampel2020system}
F.~Lampel \emph{et~al.}, ``System level synchronization of phase-coded {FMCW}
  automotive radars for {RadCom},'' in \emph{2020 14th European Conference on
  Antennas and Propagation (EuCAP)}.\hskip 1em plus 0.5em minus 0.4em\relax
  IEEE, 2020, pp. 1--5.

\bibitem{krishnamurthy2021adversarial}
V.~Krishnamurthy, K.~Pattanayak, S.~Gogineni, B.~Kang, and M.~Rangaswamy,
  ``Adversarial radar inference: Inverse tracking, identifying cognition, and
  designing smart interference,'' \emph{IEEE Transactions on Aerospace and
  Electronic Systems}, vol.~57, no.~4, pp. 2067--2081, 2021.

\bibitem{brown2012introKf}
R.~G. Brown and P.~Y. Hwang, \emph{Introduction to Random Signals and Applied
  Kalman Filtering}.\hskip 1em plus 0.5em minus 0.4em\relax Wiley, 2012.

\bibitem{mucci1984comparison}
R.~Mucci, ``A comparison of efficient beamforming algorithms,'' \emph{IEEE
  Transactions on Acoustics, Speech, and Signal Processing}, vol.~32, no.~3,
  pp. 548--558, 1984.

\bibitem{radarsp_richards}
M.~A. Richards, \emph{Fundamentals of Radar Signal Processing}, 2nd~ed.\hskip
  1em plus 0.5em minus 0.4em\relax New York: McGraw-Hill, 2014.

\bibitem{tsin2004correlation}
Y.~Tsin and T.~Kanade, ``A correlation-based approach to robust point set
  registration,'' in \emph{European conference on computer vision}.\hskip 1em
  plus 0.5em minus 0.4em\relax Springer, 2004, pp. 558--569.

\bibitem{viterbi1983nonlinear}
A.~J. Viterbi and A.~M. Viterbi, ``Nonlinear estimation of {PSK}-modulated
  carrier phase with application to burst digital transmission,'' \emph{IEEE
  Transactions on Information theory}, vol.~29, no.~4, pp. 543--551, 1983.

\end{thebibliography}
	}
\end{document}